\documentclass[aps,onecolumn,amsmath,showpacs,showkeys]{revtex4}
\usepackage{graphics}
\usepackage{graphicx}
\begin{document}

\title{Magnetic Miniband Structure and Quantum Oscillations in 
Lateral Semiconductor Superlattices}

\author{M.~Langenbuch}
\email{michael.langenbuch@physik.uni-regensburg.de}
\author{M.~Suhrke} 
\author{U.~R\"{o}ssler}

\affiliation{Institut f\"{u}r Theoretische Physik - Universit\"{a}t
Regensburg, 93040~Regensburg, Germany}

\date{\today}

\begin{abstract}
We present fully quantum-mechanical magnetotransport calculations for
short-period lateral superlattices with one-dimensional
electrostatic modulation.  A non-perturbative treatment of both
magnetic field and modulation potential proves to be necessary to
reproduce novel quantum oscillations in the magnetoresistance found in
recent experiments in the resistance
component parallel to the modulation potential. In addition, we
predict oscillations of opposite phase in the component 
perpendicular to the modulation not yet observed experimentally.  We
show that the new oscillations originate from the magnetic miniband
structure in the regime of overlapping minibands.
\end{abstract}


\maketitle

\section{Introduction}
The energy band structure of electrons in a periodic potential is one
of the fundamental concepts in solid state physics. Developed in view
of natural crystalline structures and in order to understand their
electronic properties the concept of Bloch electrons applies as well
to man-made periodic structures, which have become available with the
advent of modern technological methods of growth and structuring
\cite{champ,weiss}. While in natural crystalline structures the
lattice constant is fixed by the chemical binding mechanisms it can be
varied over a wide range in superlattices produced by controlled
technological processes. This development has renewed the interest in
the challenging problem of {\em Bloch electrons in a magnetic field},
which was studied in the past by renowned theorists \cite{bloch} but is
experimentally accessible now with these artificial periodic
structures, in which the lattice period $a$ is in the same range as
the classical cyclotron radius $R_{\mathrm{c}}$ at available magnetic
fields.

Lateral superlattices are realized by applying lithographic techniques
to impose a periodic pattern onto a two-dimensional electron system
(2DES) at a semiconductor heterointerface \cite{weiss}. This pattern
can be an uni- or bi-directional potential modulation of different
strength, thus allowing for a variety of artificial periodic
structures. Magnetotransport experiments have become an efficient
method for the investigation of the energy spectrum.
At the beginning\cite{weis89_wink89}, with realizable periods $a$ of
these lateral superlattices (several $100\,$nm) much larger than the
typical Fermi wavelengths $\lambda_{\mathrm F}$ of the 2DES ({\it
e.g.}  $50\,$nm), classical concepts have been successful in
interpreting the magnetotransport data.
 One prominent example was the observation of $1/B$
periodic oscillations in the low magnetic field range which could be
ascribed to the commensurability condition $2R_{c}= a (\nu-\frac{1}{4}),
\nu=1,2,3...$ \cite{been89}. With decreasing lattice periods (at
present $a$ \raisebox{-0.5ex}{$\stackrel{<}{\sim}$} $100\,$nm)
quantum
mechanical aspects gain importance. In fact, a new type of $1/B$
periodic oscillations has been detected recently in lateral
semiconductor superlattices with bi- \cite{albrecht} and
uni-directional \cite{deutschmann} electrostatic modulation. Their
interpretation as being due to the formation of an artificial band
structure was based on semiclassical arguments involving modifications
of the Fermi contour of the 2DES by the superlattice potential and
magnetic breakdown. It was possible to quantitatively relate the
period of these novel quantum oscillations (including their dependence
on electron density $n_{\mathrm s}$ and superlattice constant $a$) to the area enclosed by the Fermi contours
\cite{albrecht,deutschmann,lgb00}. 

However, except for the magnetic
breakdown this explanation did not address the combined effect of
periodic potential and magnetic field that gives rise to a
magnetic miniband structure.

In this letter we present fully quantum-mechanical magnetotransport
calculations by evaluating Kubo's formula for a 2DES with non-perturbative treatment of both
uni-directional periodic potential and magnetic field. It is shown
that the new $1/B$ periodic quantum oscillations reported in
\cite{albrecht,deutschmann} result from the formation of magnetic
minibands which (at fixed magnetic field) evolve with increasing
amplitude of the potential modulation from the Landau levels of the
2DES. Besides the
impurity scattering, here considered in the self-consistent Born
approximation (SCBA), it is the dispersion of these magnetic minibands, that determines the magnetoconductivity. As 
will be shown, both mechanisms enter differently into the
conductivities parallel and perpendicular to the potential modulation
thus causing a phase shift by $\pi$ in the novel quantum oscillations.
The experimental detection of this phase shift together with the
period of these magnetotransport oscillations can serve as evidence
for the formation of magnetic minibands.

\section{Magnetic Miniband Structure}
We consider electrons with effective mass $m^{\ast}$ moving in the
xy-plane subjected to a periodic electrostatic potential $V(x)=
\frac{V_{0}}{2} \cos \left( \frac{2 \pi}{a} x \right)$. The
single-particle Hamiltonian is \cite{gerh89,vasi89,zhan90}
\begin{equation}
\label{eqn_hamilton}
H = \frac{1}{2 m^{\ast}} \left( {\mathbf{p}} + e {\mathbf{A}}
\right)^2 + V(x),
\end{equation}
where the vector potential $\mathbf{A}$ includes the magnetic field in
z-direction and is given in Landau gauge by ${\mathbf A}(x) = Bx
\hat{\mathbf e }_{y}$. The amplitude $V_0$ of the potential modulation
will be varied with respect to the Fermi energy $E_{\mathrm{F}}$ of the
2DES from weak ($V_0\ll E_{\mathrm {F}}$) to strong
($V_0$\raisebox{-0.5ex}{$\stackrel{<}{\sim}$}$E_{\mathrm {F}}$). Due
to the Landau gauge the Hamiltonian commutes with the $y$-component of
the momentum.  Without lateral superlattice ($V_0=0$) the eigenstates
can be characterized by two quantum numbers, the Landau level index
$n_{\mathrm {L}}$ and the center coordinate $X_{0}$, which by $X_{0} =
- \lambda_{\mathrm c}^2 k_{y}$ is connected with the momentum $\hbar
k_y$ in $y$-direction ($\lambda_{\mathrm c}^2=\hbar/eB$).  In real
space representation the eigenstates $|n_{\mathrm L} X_{0}\rangle$ are
given by
\begin{equation}
\Psi_{n_{\mathrm{L}}X_{0}}(x,y) =
\varphi_{n_{\mathrm{L}}} (x-X_{0}) \frac{1}{\sqrt{L_{y}}} e^{ik_{y}y}
\end{equation}
where $\varphi_{n_{\mathrm L}}(\xi)$ is the eigenfunction of the
harmonic oscillator and the energy eigenvalues $E_{n_{\mathrm L}} =
\hbar \omega_{\mathrm c} (n_{\mathrm L} + \frac{1}{2}), n_{\mathrm L}
= 0,1,2,\ldots$ depend only on the Landau level index. $\omega_{c}=\frac{e B}{m^{*}}$ is the cyclotron frequency. The spatial
homogeneity of the unmodulated 2DES leads to a degeneracy of the
Landau levels given by the ratio $\Phi/\Phi_0$, of the magnetic flux
$\Phi$ through the sample area $L_xL_y$ and the elementary flux
quantum $\Phi_0 = h/e$. When considering the superlattice potential
this degeneracy is lifted and the Landau levels evolve into magnetic
minibands described by a dependence of the energy eigenvalues on
$X_0$(or $k_y$), which due to the periodicity of the potential can be
restricted to the interval $[-\frac{a}{2},\frac{a}{2}]$.
 
\begin{figure}
\includegraphics[scale=0.9]{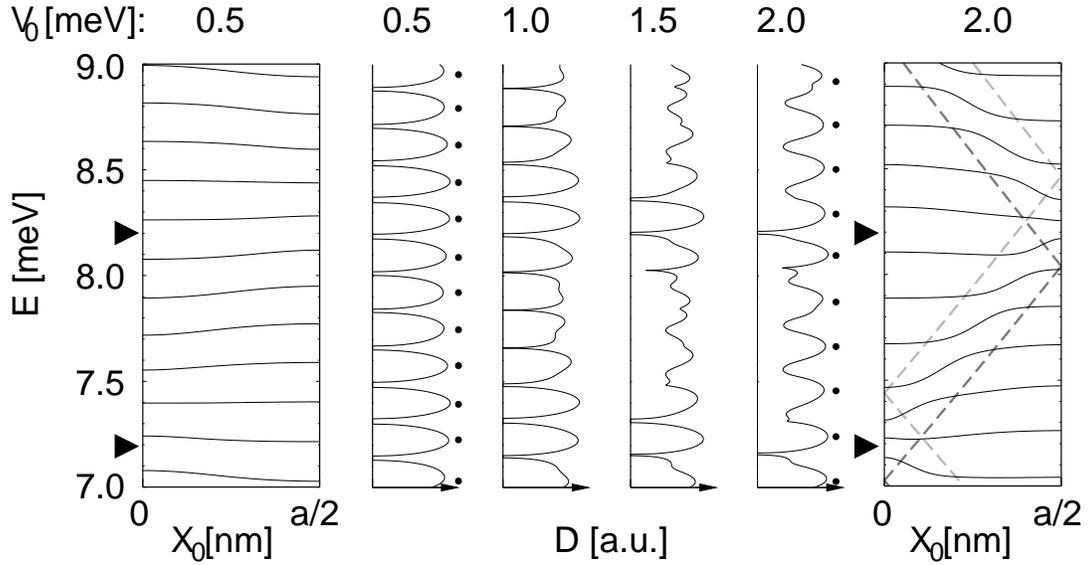} 
\caption{\label{fig_bands} 
Part of the magnetic miniband structure (left and right) and density
of states $D(E)$ (center) for uni-directional modulation with
$a=100\,$nm, $\mu = 50\,$ m$^2$/Vs, $m^{\ast}= 0.067 \, m_{0}$ and
different modulation amplitudes $V_{0}$ at a magnetic field $B =
0.10\,$T. The maxima of the density of states are marked for $V_{0} =
0.5\,$meV and $V_{0} = 2.0\,$meV.  The dashed lines in the 
miniband structure for $V_{0}=2.0\,$meV is the dispersion of free (light grey) and
constantly Bragg reflected electrons (dark grey) [15]. The triangles at the
energy axes mark the flat-band conditions.}
\end{figure}

The solution of the eigenvalue problem with $H$ from
Eq.~(\ref{eqn_hamilton}) is the basis for the calculation of the
magnetoconductivity. (In our calculations we have used the effective
mass $m^{\ast}=0.067\, m_0$, typical for electrons in GaAs, the
lattice constant $a=100\,$nm, and the carrier density n$_{s} = 2 \cdot
10^{15}\,$m$^{-2}$ corresponding to $E_{\mathrm F}= 7.1\,$meV).  The
energy spectrum at fixed magnetic field ($B=0.1\,$T) is shown in
Fig.~\ref{fig_bands} for two different strengths of the potential
modulation $V_0$ (left and right part).

For weak modulation ($V_0=0.5\,$meV) the Landau level structure can
still be recognized, but the degeneracy is lifted already and gives
rise to magnetic minibands with weak dispersion. With increasing $V_0$
the dispersion gets stronger and the magnetic minibands overlap, thus
the Landau level index looses its meaning and has to be replaced by
the magnetic miniband index $n$. A striking feature of the magnetic
miniband structure for strong modulation ($V_{0} = 2\,$meV) are the
sections with pronounced dispersion. Their classical analogue are
channeling and drifting orbits \cite{zwer99}. These sections fall into
limits defined by the dispersion of free and constantly Bragg
reflected electrons \cite{fussnote}.

In previous work \cite{gerh89,vasi89,zhan90} the potential modulation
was treated as first order perturbation correction to the Landau
levels.  This treatment is justified if the potential modulation is
much smaller than $\hbar\omega_c$ and the minibands (Landau bands)
still reflect the Landau level structure. This approach provided a
quantum mechanical explanation of the commensurability oscillations as
being due to the flat-band condition indicated by triangles in
Fig.~\ref{fig_bands}. In the quantum mechanical approach this
condition can be formulated as $R_{c} = \lambda_{c}^2 k $, with $k$
taken from the free electron dispersion relation $E(k)=\frac{\hbar^{2}
k^{2}}{2 m^{\ast}}$ and coincides with the maxima of the
DOS-envelope. In this approach also the observed antiphase
commensurability oscillations in the two diagonal resistivity components could be
explained by invoking the two transport mechanisms due to Landau band
dispersion and impurity scattering. However, first order perturbation theory is not adequate to
describe the magnetotransport at very low magnetic fields, when the
superlattice potential mixes the Landau levels into magnetic
minibands.

\section{Conductivity}
Following a standard routine \cite{zhan90} we calculate the
conductivity tensor in linear response theory by evaluating the Kubo
formula \cite{stre75,kubo57}:
\begin{equation}
\label{eqn_cond}
 \sigma_{\mu \nu} = \frac{i e^2 \hbar}{L_{x} L_{y}}
\int dE f_{0}(E)  \left \langle \mbox{Sp} 
\left\{  v_{\mu} \frac{d{\cal G}^{+}}{dE} v_{\nu} \delta(E\!-\!H)
- v_{\mu} \delta(E\!-\!H) v_{\nu} \frac{d{\cal G}^{-}}{dE}
\right\}
\right \rangle_{\mathrm I},
\end{equation}
where $v_{\nu}$ are the components of the velocity operator $\mu,\nu
\in \{x,y\}$, and ${\cal G}^{\pm}$ is the retarded ($+$) and advanced ($-$) Green
function. The Fermi distribution $f_{0}(E)$ includes the temperature
and $\langle ...\rangle_{\mathrm I}$ denotes an average over impurity
configurations.
This average is performed in SCBA by assuming short range
scatterers. In the complex number approximation it yields a
single-particle self-energy given by
\begin{equation}
\label{eqn_sigma}
\Sigma^{+}(E)  = \Gamma^2 \sum_{n}  \frac{1}{a} \int_{-a/2}^{a/2} dX_{0}
 G_{n X_{0}}^{+}(E)\,,
\end{equation} 
where $\Gamma^2 = \frac{1}{2 \pi} \hbar \omega_{c}
\frac{\hbar}{\tau}$ is connected through the relaxation time $\tau$ with the zero field mobility $\mu =
e \tau/m^{\ast}$. The impurity average in Eq.~(\ref{eqn_cond}) is
considered by replacing ${\cal G}^{\pm}$ by the impurity averaged Green
functions $G^{\pm} = \langle {\cal G}^{\pm}\rangle_{\mathrm I}$. Vertex corrections,
which vanish in this approximation for the homogeneous 2DES, are
neglected here.  In the following we concentrate on the longitudinal
conductivity $\sigma_{\mu \mu}$ which can be written as:
\begin{equation}
\label{eqn_sigmalong}
\sigma_{\mu \mu} = \frac{e^2}{h} \frac{\hbar^2 \pi}{\lambda_{\mathrm c}^2} \int dE 
\left( - \frac{df_{0}(E)}{dE} \right)
\sum_{n_{1},n_{2}} \frac{1}{a} 
\int_{-a/2}^{a/2}  dX_{0} 
\left| \left< n_{1} X_{0} \left| v_{\mu} \right| n_{2} X_{0} \right>
\right|^{2} A_{n_{1} X_{0}}  A_{n_{2}X_{0}} , 
\end{equation}
where $n_{1}$, $n_{2}$ denotes the magnetic miniband index. The
spectral functions \linebreak[4] $A(E) = \frac{i}{2 \pi} \left(
G^{+}(E) - G^{-}(E) \right)$ are obtained from the impurity averaged
Green functions.  In SCBA the density of states (DOS) can be expressed
by:
\begin{equation}
\label{eqn_dos} 
D(E) = \frac{1}{L_{x}L_{y}}\, \mbox{tr}\{A(E)\} = - \frac{1}{\left(
\pi \Gamma \lambda_{c} \right)^{2}}
\,\mbox{Im} \left\{ \Sigma^{+}(E) \right\}. 
\end{equation}
The center part of Fig.~\ref{fig_bands} shows how $D(E)$ evolves with
increasing strength $V_{0}$ of the potential modulation. Weak
modulation ($V_0=0.5\,$meV) causes only a slight dispersion and the
DOS with well separated peaks is still reminiscent of the Landau level
structure of the homogeneous 2DES as indicated by dots. This
characterizes the high-field regime of magnetotransport with the
Shubnikov-de Haas (SdH) oscillations. With increasing modulation
amplitude $V_0$ the energy spectrum gets more complex until due to the
strong coupling of Landau levels the DOS exhibits a new periodicity
(indicated by the dots in the panel for $V_0=2.0\,$meV).

In a simple picture, neglecting in Eq.~(\ref{eqn_sigmalong}) the
dependence of the matrixelement on quantum numbers, the conductivity
is related to the DOS expressed by Eq.~(\ref{eqn_dos}). Thus it is
conceivable from the discussion of Fig.~\ref{fig_bands} that due to
formation of magnetic minibands with strong dispersion novel $1/B$
periodic quantum oscillations can show up with periods differing from
those of commensurability and SdH oscillations.

\begin{figure}

\begin{center}
\includegraphics[scale=0.7]{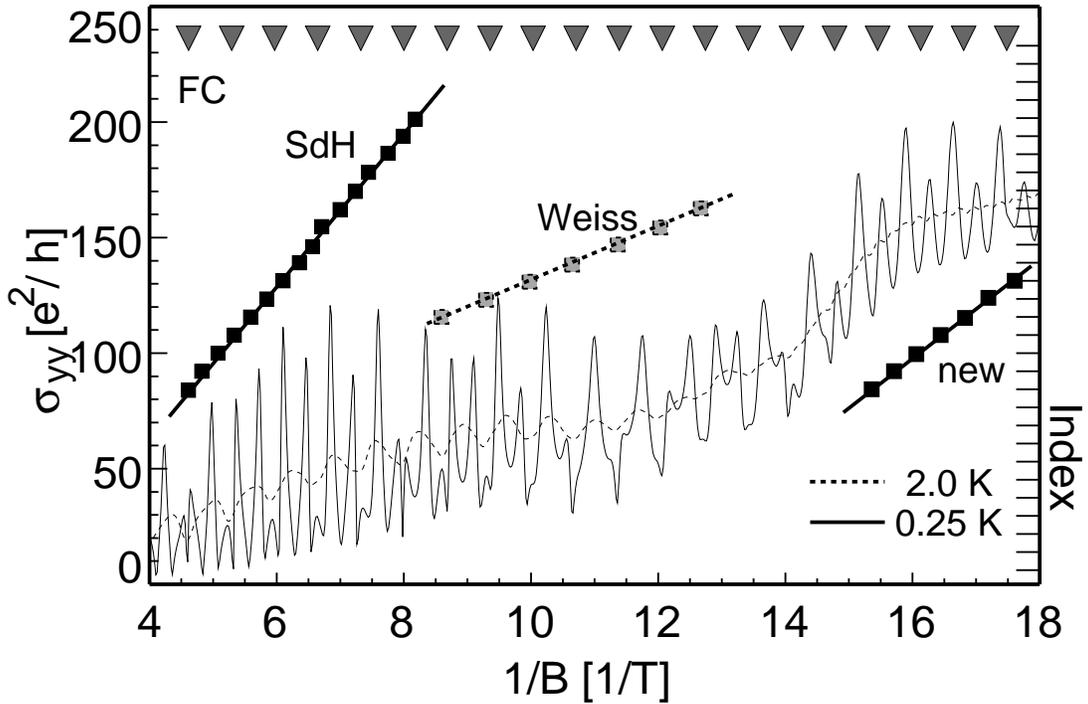} 
\end{center}
\caption{\label{fig_two} Conductivity $\sigma_{yy}$ perpendicular to
the modulation direction  at $T=2.0 \,{\rm K}$ (dotted curve) and
$0.25\,$K (solid curve) for an uni-directional modulation with $V_{0} = 2.0\,$meV and $a=100\,$nm. The electron system is charcterized by  $m^{\ast}= 0.067 \, m_{0}$, $\mu = 50\,$ m$^2$/Vs and $n_{s}
= 2.0 \cdot 10^{15}\,$m$^{-2}$. The flat-band conditions are marked by
triangles.  The $1/B$ periodicities of SdH $\Delta_{1/B} =
0.24\,$T$^{-1}$, commensurability (Weiss)
$\Delta_{1/B} = 0.68\,$T$^{-1}$ and new quantum oscillations  $\Delta_{1/B} =
0.37\,$T$^{-1}$ are indicated, by assigning an index to the position of the minima in the calculated conductivity. For the commensurability oscillations the minima at $T=2.0\,$K are taken.}
\end{figure}

\begin{figure}

\begin{center}
\includegraphics[scale=0.7]{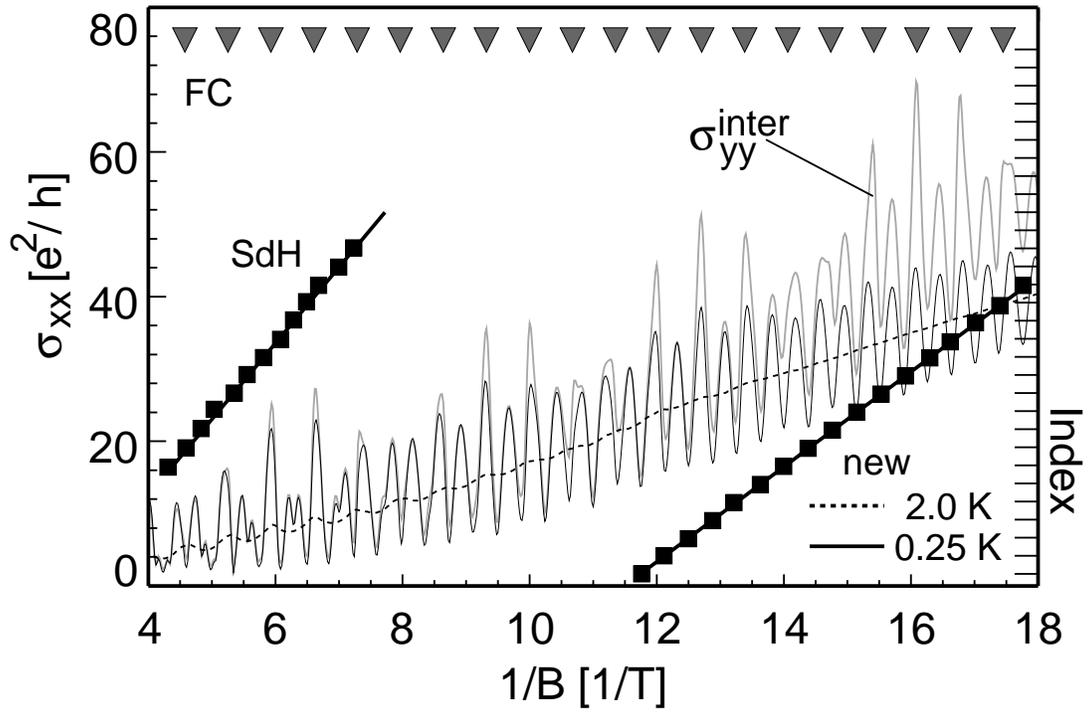}
\end{center}
\caption{\label{fig_one} Same as Fig.~\ref{fig_two} but conductivity $\sigma_{xx}=\sigma_{xx}^{\mathrm{inter}}$ parallel to the modulation is shown. For comparison we plot 
$\sigma_{yy}^{\mathrm{inter}}$, the inter-miniband contribution to
$\sigma_{yy}$.  The flat-band conditions (FC) are marked by triangles. The $1/B$ periodicities of SdH oscillations  and of the new quantum oscillations (new) are indicated.}
\end{figure}

In order to confirm this preliminary conclusion we turn now to the
calculated conductivities perpendicular
($\sigma_{yy}$) and parallel ($\sigma_{xx}$)  to the potential modulation as shown in
Figs.~\ref{fig_two} and \ref{fig_one}, respectively.

The conductivities (for two different temperatures) are plotted here
versus $1/B$ to expose the characteristic periodicities. In the
high field regime, when the potential modulation is much weaker than
$\hbar\omega_c$, the conductivity is determined by the SdH
oscillations characteristic for the Landau level spectrum.  In the
intermediate field regime the periods of the Weiss oscillations,
determined by the commensurability condition, are marked for
$\sigma_{yy}$ (Fig.~\ref{fig_two}).

\begin{figure}
\begin{center}
\includegraphics[scale=0.7]{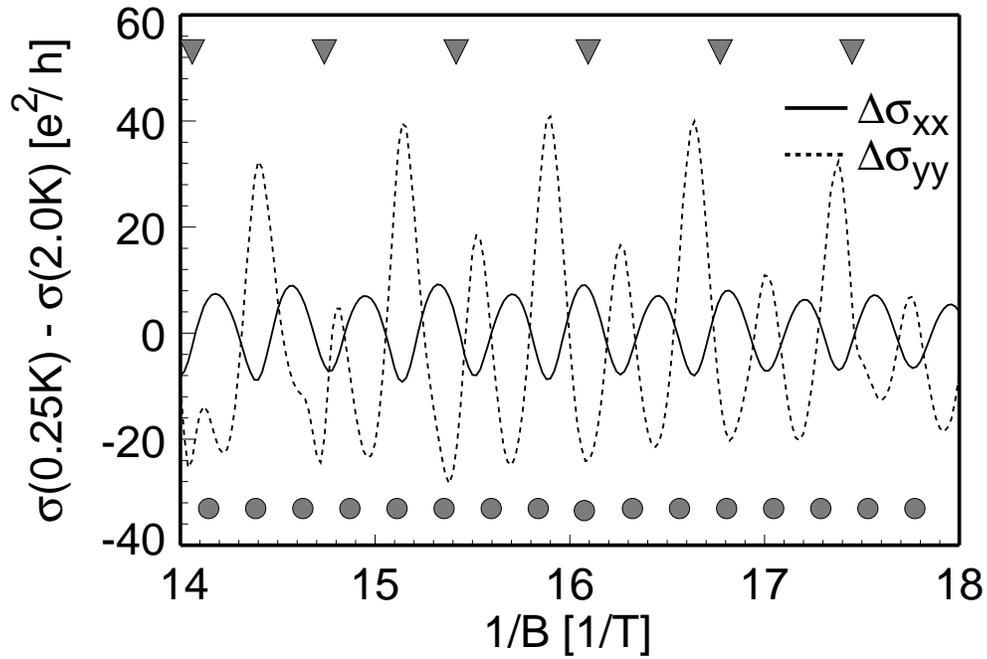} 
\end{center}
\caption{\label{fig_three} Difference of the longitudinal conductivities at $0.25\,$K and 
$2.0\,$K with the parameters of Fig.~\ref{fig_two}. The flat-band
conditions are marked by triangles and the expected SdH maxima by
grey bullets.}
\end{figure}

Oscillations with a new period $\Delta_{1/B} = 0.37\,$T$^{-1}$ show up
in both conductivity components in the low-field regime. It is
characteristic for the new oscillations to vanish by increasing the
temperature from $0.25\,$K (solid curves) to $2.0\,$K (dotted curves),
while the Weiss oscillations can still be seen.  In order to
understand these results we consult
Eq.~(\ref{eqn_sigmalong}). Contributions with $n_{1} \neq n_{2}$ are
determined by the overlapping spectral functions from different magnetic minibands
and will be called inter-miniband conductivity
$\sigma^{\mathrm{inter}}$. For weak modulation, when the overlap is caused exclusively by the SCBA broadening alone, this contribution
can be ascribed to impurity scattering and is also called scattering
conductivity \cite{gerh89,vasi89,zhan90}. The second mechanism
deriving from terms with $n_{1} = n_{2}$ is due to the dispersion of
the magnetic minibands exclusively (the matrixelements in
Eq.~(\ref{eqn_sigmalong}) correspond to the group velocity) and will
be called intra-band conductivity $\sigma^{\mathrm{intra}}$.  For
systems with uni-directional potential modulation in $x$ direction the
matrixelement $\left< n X_{0} \left| v_{x} \right| n X_{0} \right>$
vanishes and the intra-band conductivity contributes only to
$\sigma_{yy}$. Thus the use of 1D lateral superlattices permits the observation of quantum oscillations of scattering and band contributions to the conductivity at the same time. Note that up to present quantum oscillations in the 1D case \cite{deutschmann} have been found only in the magnetoresistance parallel to the modulation which corresponds to the perpendicular conductivity. As can be seen in Fig.~\ref{fig_one}, $\sigma_{xx}$ is almost identical (in
magnitude and phase) with $\sigma_{yy}^{\mathrm{inter}}$ except for the low
field regime where the new oscillations occur. We focus our attention
to this regime and to the new oscillations by showing in
Fig.~\ref{fig_three} the difference between the conductivities
calculated for the two temperatures.
As can be seen clearly $\Delta \sigma_{xx} = \sigma_{xx}(0.25\,
{\mathrm K}) - \sigma_{xx} (2.0\, {\mathrm K})$ and $\Delta
\sigma_{yy}$ oscillate in antiphase. This is a consequence of the two
mechanisms (inter- and intra- miniband) contributing differently to
the conductivity parallel and perpendicular to the potential
modulation and thus reflects the formation of magnetic minibands:
$\sigma_{xx}$ is dominated by inter-miniband contributions with maxima
correlated with those of the DOS; $\sigma_{yy}$ is dominated (for the
parameters used here) by intra-miniband contributions, that get large
when parts of the magnetic miniband with strong dispersion, which
correspond to small DOS, pass through the Fermi energy. A similar
explanation has been invoked to explain the observed antiphase in
commensurability oscillations yet under the assumption of weak
modulation and only slightly distorted Landau band spectrum
\cite{gerh89,vasi89,zhan90}. Here we provide more rigorous analysis
for arbitrary modulation strength (or low magnetic field) at which the
new $1/B$ periodic oscillations appear which together with the
antiphase signature, if observed experimentally, can be taken as a
proof for the formation of the magnetic minibands in 2DES with
uni-directional potential modulation.  In general, as the oscillations
have been observed for 2D modulated systems too \cite{albrecht}, the
periodicity of the quantum oscillations can be explained by the
density of states but the phase depends on the dominating conductivity
mechanism.

\section{Summary}
In conclusion we address one of the challenging single-particle 
problems in solid state physics: Bloch electrons in a magnetic field
and a possible proof of the formation of a magnetic band
structure. For this purpose we perform a quantum-mechanical
calculation of the magneto conductivity by evaluating the Kubo formula
for a system of 2D electrons with an uni-directional periodic potential
modulation in an external homogeneous magnetic field included non-perturbatively. Such systems are
realized nowadays in layered semiconductor structures with modern
technologies of growth and lithography. It is shown that the formation
of a magnetic miniband structure together with the impurity scattering
manifests in the magneto conductivities parallel and perpendicular to
the potential modulation: in the low magnetic field regime new $1/B$
periodic quantum oscillations show up, which are in antiphase for the
two longitudinal components of the conductivity tensor due to different conductivity mechanisms and should be
observable only at very low temperatures.

This work has been supported by the DFG via Forschergruppe~370 {\em
Ferromagnet-Halbleiter-Nanostrukturen}.

\clearpage
\newpage

\begin{thebibliography}{9}
\bibitem{fussnote} Due to the modulation so called channeling orbits exist whose energies
are not quantized in the magnetic field.  By
avoiding bandcrossings the continuous energy
dispersion and the discrete energies of the Landau ladder form the magnetic miniband structure for high
$V_{0}$. The energy dispersions of
electron waves moving straight in y-direction, $E(k) =
\frac{\hbar^2 k^2}{2 m^{\ast}}$, and of wave packets consisting of electrons 
which are Bragg reflected at $k_{x} = \frac{\pi}{a}$, $E(k) =
\frac{\hbar^2 k^2}{2 m^{\ast}} -\frac{1}{12} \frac{\hbar^{2}
\pi^2}{ m^{\ast} a} \ln (1+6 \frac{k^2 a^2}{\pi^2})$,  are marked  in Fig.~\ref{fig_bands}(right). These limits separate the dispersive part of the
magnetic minibands from the flat one  for our parameters. Note that the Bragg reflected wave is  no classical path in contrast to the trajectories considered in \cite{zwer99} to explain the miniband structure.
\end{thebibliography}
\end{document}